\documentclass[10pt,twoside,BCOR7mm,DIV15,headinclude,footexclude,
               cleardoubleempty,idxtotoc]
{scrartcl}

\usepackage{natbib}
\usepackage[font=small,labelfont=bf]{caption}
\usepackage[english]{babel}
\usepackage{graphicx}
\usepackage{hyperref}
\usepackage{scrpage2}
\usepackage{ifthen}
\usepackage{booktabs}
\usepackage{amsmath}
\usepackage{amssymb}
\usepackage{multicol}
\usepackage{float}
\usepackage{hyperref}

\hypersetup{breaklinks=true
,colorlinks=true,linkcolor=black,urlcolor=blue
,citecolor=black}

\addto\captionsenglish{%
}

\makeatletter
\renewcommand{\@biblabel}[1]{}
\renewcommand{\@cite}[2]{%
{#1\ifthenelse{\boolean{@tempswa}}{,#2}{}}}
\makeatother
\ofoot{\thepage}
\ifoot{}

\setcapindent{0em}
\setheadsepline{1pt}

\setkomafont{pagehead}{\normalfont\sffamily}
\setkomafont{pagenumber}{\normalfont\rmfamily}

\makeatletter
\newcommand{\listofcontributions}{\@starttoc{con}}

\newcommand{\l@contribution} {\@dottedtocline{1}{1.5em}{2.3em}}
\makeatother

\newenvironment{contribution}{
\setcounter{section}{0}
\setcounter{figure}{0}
\setcounter{table}{0}
}{
\newpage
\lehead{}
\rohead{}
}

\begin{document}

\setlength{\baselineskip}{2.5ex}

\begin{contribution}

\lehead{J.A.\,Toal\'{a} et al.}

\rohead{Diffuse X-ray Emission within WR Nebulae}

\begin{center}
{\LARGE \bf Diffuse X-ray Emission within Wolf-Rayet Nebulae}\\
\medskip

{\it\bf J.A.\,Toal\'{a}$^1$, M.A.\,Guerrero$^1$, Y.-H.\,Chu$^{2}$, S.J.\,Arthur$^{3}$ \& R.A.\,Gruendl$^4$}\\

{\it $^1$Instituto de Astrof\'{i}sica de Andaluc\'{i}a, IAA-CSIC, Granada, Spain}\\
{\it $^2$Institute of Astronomy and Astrophysics, Academia Sinica (ASIAA), Taipei, Taiwan}\\
{\it $^3$Centro de Radioastronom\'{i}a y Astrof\'{i}sica, UNAM, Campus Morelia, Mexico}\\
{\it $^4$Department of Astronomy, University of Illinois, Urbana, IL, USA}

\begin{abstract}

  We discuss our most recent findings on the diffuse X-ray emission
  from Wolf-Rayet (WR) nebulae. The best-quality X-ray observations of
  these objects are those performed by {\it XMM-Newton} and {\it
    Chandra} towards S\,308, NGC\,2359, and NGC\,6888. Even though
  these three WR nebulae might have different formation scenarios,
  they all share similar characteristics: i) the main plasma
  temperatures of the X-ray-emitting gas is found to be
  $T$=[1--2]$\times$10$^{6}$~K, ii) the diffuse X-ray emission is
  confined inside the [O\,{\sc iii}] shell, and iii) their X-ray
  luminosities and electron densities in the 0.3--2.0~keV energy range
  are $L_\mathrm{X}\approx$10$^{33}$--10$^{34}$~erg~s$^{-1}$ and
  $n_\mathrm{e}\approx$0.1--1~cm$^{-3}$, respectively. These
  properties and the nebular-like abundances of the hot gas suggest
  mixing and/or thermal conduction is taking an important r\^{o}le
  reducing the temperature of the hot bubble.

\end{abstract}
\end{center}

\begin{multicols}{2}

\section{Introduction}

Massive stars represent the main source of feedback that govern the
physics of the Interstellar Medium (ISM). 
The most massive stars ($M_\mathrm{i}\gtrsim$25~M$_{\odot}$) will
evolve through the red supergiant or yellow supergiant (RSG or YSG) or
luminous blue variable (LBV) phase depositing up to half their masses
into the ISM, 
to finally become Wolf-Rayet (WR) stars
\citep[e.g.,][and references therein]{Ek2012}.
During the intermediate RSG/YSG or LBV phase, the star develops a slow
and dense wind ($\dot{M}$=10$^{-4}$--10$^{-3}$~M$_{\odot}$~yr$^{-1}$,
$v_{\infty}$=10--100~km~s$^{-1}$) with no significant UV flux. The
final WR phase is characterized by a strong wind
($\dot{M}$=10$^{-5}$~M$_{\odot}$~yr$^{-1}$,
$v_{\infty}$=1500~km~s$^{-1}$) that sweeps up, shocks, and compresses
the RSG/LBV slow material, while a newly developed ionising photon
flux ionises the material. This combination of effects will lead to
the formation of the so-called WR nebulae (or ring nebulae).


Before the current generation of X-ray satellites (e.g., {\it
  XMM-Newton}, {\it Chandra}, and {\it Suzaku}) the only WR nebulae
reported to harbor diffuse X-ray emission were S\,308 and NGC\,6888
around WR\,6 and WR\,136, respectively \citep[see,
e.g.,][]{Wrigge1999,Wrigge2005} but with low resolution X-ray
maps. The reported plasma temperatures and X-ray luminosities of the
X-ray-emitting gas were $T\approx$10$^{6}$~K and
$L_\mathrm{X}\approx$10$^{34}$~erg~s$^{-1}$, respectively.

In this talk we review the most recent {\it Chandra} and {\it
  XMM-Newton} observations towards S\,308, NGC\,2359, and NGC\,6888
  around WR\,6, WR\,7 and WR\,136, respectively.




\section{On the origin of the diffuse X-ray emission}

An adiabatically shocked stellar wind can form a hot bubble inside a
nebula with an estimated temperature of $k_\mathrm{B}T = 3 \mu
m_\mathrm{H} v_{\infty}^{2} /16$ \citep[see][]{Dyson1997}, where $\mu$
is the mean particle mass for fully ionised gas, $m_\mathrm{H}$ is the
hydrogen mass, and $k_\mathrm{B}$ is the Boltzmann's constant. That
is, for the estimated stellar wind velocities of WR\,6 and WR\,136 of
$v_{\infty}\approx$1700~km~s$^{-1}$ \citep[][]{Hamann2006}, hot
bubbles with temperatures of $T\sim$10$^{7}$--10$^{8}$~K are expected,
which are in clear mismatch with the observed X-ray temperatures.

It has been argued that thermal conduction between the outer cold
(10$^{4}$~K) nebular material and the hot bubble would lead to
temperatures as those as reported by X-ray observations. In
particular, the often cited work of \citet{Weaver1977}, which assumes
classical thermal conduction \citep[e.g.,][]{Spitzer1962}, predicts
higher luminosity values than those observed ($L_\mathrm{X}
\geqslant$10$^{35}$~erg~s$^{-1}$). Recent numerical studies have shed
light into this problem showing that as a result of the
slow-wind/fast-wind interaction, the swept up shell will break up due
to the formation of Rayleigh-Taylor and thin shell instabilities and
will be a source of mass, reducing the temperature of the hot bubble,
raising the density to observed values, and reproducing the observed
X-ray luminosities \citep{Toala2011,Dw2013}.


\section{Notes on individual objects}

To date, five WR nebulae have been reported in the literature to be
observed with the current generation of X-ray satellites: S\,308,
NGC\,2359, RCW\,58, and NGC\,6888 around WR\,6, WR\,7, WR\,40, and
WR\,136, respectively, and the WR nebula around WR\,16, but diffuse
X-ray emission has been detected only in S\,308, NGC\,2359, and
NGC\,6888 (see Figure~1). In this section we will present a summary of
the best-quality X-ray observations towards these three objects as
performed with {\it XMM-Newton} and {\it Chandra} satellites.

\begin{figure*}[!t]
\begin{center}
\includegraphics[width=1.9\columnwidth]{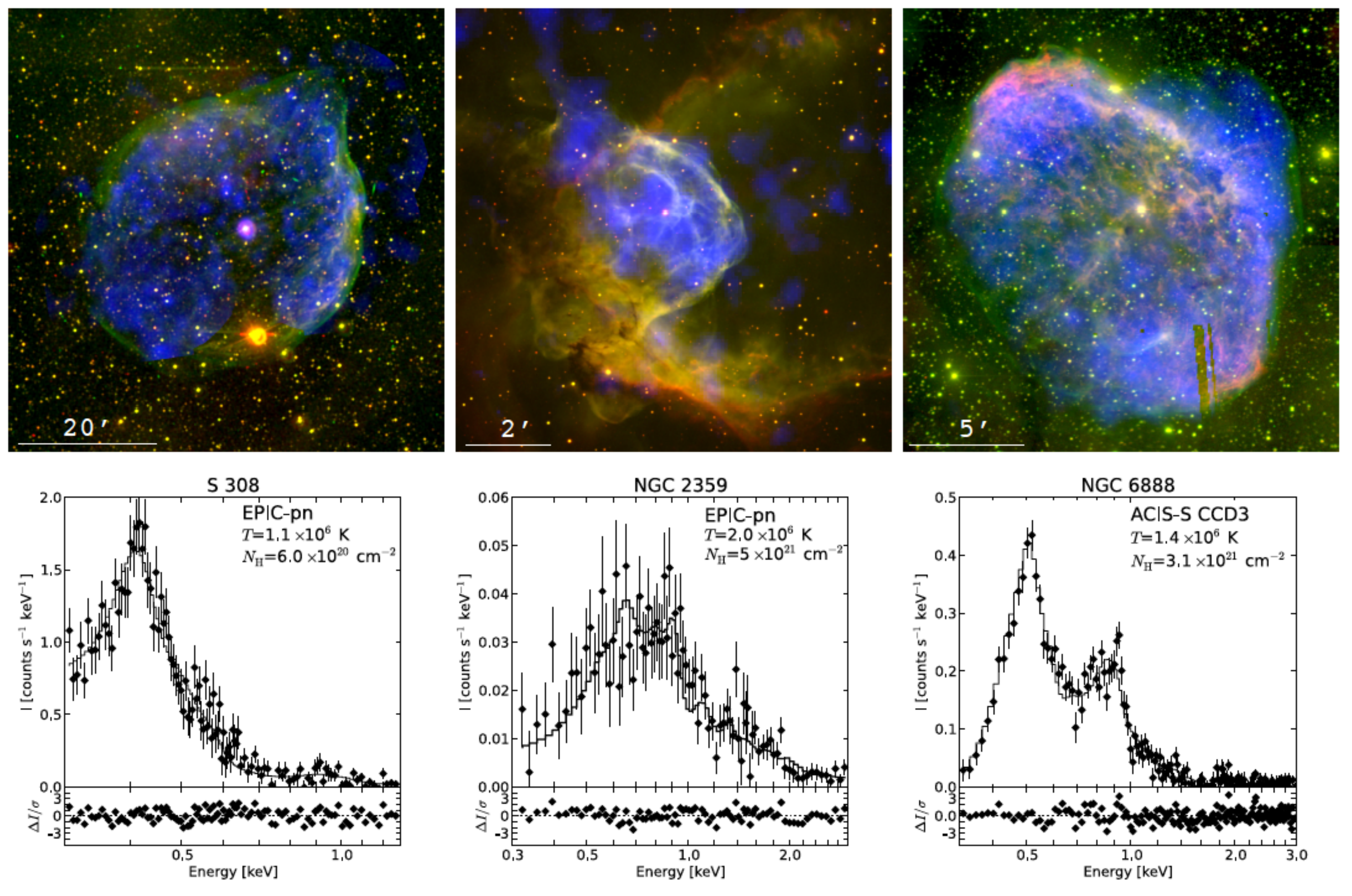}
\caption{X-ray emission from the WR nebulae S\,308 (left panels),
  NGC\,2359 (central panels), and NGC\,6888 (right panels) around WR\,6,
  WR\,7 and WR\,136, respectively. Top panels: Colour-composite
  pictures. Red, green, and blue correspond to the H$\alpha$, [O\,{\sc
    iii}], and diffuse X-ray emission as detected by {\it
    XMM-Newton}. Bottom panels: Background-subtracted spectra of the
  diffuse X-ray emission overplotted with their best-fit optically
  thin plasma model (solid line).}
\label{example:smallfig}
\end{center}
\end{figure*}

\subsection{S\,308 (WR\,6)}

S\,308 around WR\,6 is a nearly spherical nebula with an angular size
of $\sim$40$'$ (in diameter) rendering it the most extended object of
this class. Because of this, it had to be observed with 4 {\it
XMM-Newton} pointings that mapped 90\% of its total extension. The
diffuse X-ray emission was found to be confined by the [O\,{\sc iii}]
emission and displays a clear limb-brightened morphology \citep[][see
Figure~1 - upper left panel]{Toala2012}. The EPIC-pn
background-subtracted spectrum presented in Fig.~1-bottom left panel
shows that it is mainly dominated by the He-like triplet of N\,{\sc
vi} at 0.43~keV declining towards higher energies.

\citet{Toala2012} presented a spectral study of S\,308 concluding that
spectra extracted from different regions can be modeled by an
optically thin plasma model with nebular abundances and a main
temperature of $T$=1.1$\times$10$^{6}$~K. The averaged luminosity and
rms electron density are
$L_\mathrm{X}$=2$\times$10$^{33}$~erg~s$^{-1}$ and
$n_\mathrm{e}$=0.1~cm$^{-3}$.

\citet{Toala2011} argued that the most probable scenario formation of
this nebula is that the central star, WR\,6, might have evolved
through a YSG phase in order to create such an extended nebula
($\sim$9~pc in radius) using a stellar evolution model with an initial
mass of 40~M$_{\odot}$ and initial rotation of 300~km~s$^{-1}$
\citep{MM2003}.

\subsection{NGC\,2359 (WR\,7)}

The WR nebula NGC\,2359 presents an interesting morphology: it
displays a main central bubble with several blisters and filaments
(see Fig.~1, upper central panel). The X-ray-emitting gas seems to
fill the main cavity and the northeast blister. It is probable that
this is also the case for the southeast blister, but molecular
material in the line of sight towards this region precludes a clear
view of the total distribution of the X-ray-emitting gas
\citep[see][and references therein]{Rizzo2003}. Furthermore, as different 
velocity components have been reported by \citet{Rizzo2003} and the
complex shape of the nebula points out an eruptive and non-isotropic
origin as an LBV \citep{Toala2015b}.

The EPIC-pn background-subtracted spectrum presents a broad peak
around 0.5--0.9~keV, with two apparent maxima at 0.65~keV and
$\sim$0.9~keV (Fig.~1, bottom central panel). \citet{Toala2015b}
estimated a main temperature of $T$=2$\times$10$^{6}$~K adopting
nebular abundances. The luminosity and rms electron density were
estimated to be $L_\mathrm{X}$=2$\times$10$^{33}$~erg~s$^{-1}$ and
$n_\mathrm{e}\lesssim$0.6~cm$^{-3}$.

\subsection{NGC\,6888 (WR\,136)}

NGC\,6888 is the most studied WR nebula in optical and X-rays. Its
H$\alpha$ emission shows a nearly elliptical distribution of clumps
but the [O\,{\sc iii}] emission reveals a more spherical morphology
with a blowout towards the northwest (Fig.~1, upper right
panel). 
\citet{Toala2014} reported the study of the spectral properties as
observed with {\it Chandra} ACIS-S CCD\#3 and \#4. Even though the
spectral responses of these CCDs are not the same and the observations
only covered 60\% of the nebula, they manage to derive the physical
parameters of the X-ray-emitting gas. \citet{Toala2014} argued that
there should be an extra maximum in the spatial distribution of the
X-ray emission towards the northwest blowout.

The {\it Chandra} ACIS-S CCD\#3 spectrum shows that there are two main
components, one at the N\,{\sc vii} at 0.5~keV and a secondary peak at
0.7--0.9~keV which could be associated to the Fe complex and Ne lines
(see Fig.~1, bottom right panel). The estimated main plasma
temperature was $T$=1.4$\times$10$^{6}$~K assuming an optically thin
plasma model with nebular abundances. The luminosity and rms electron
density were estimated to be
$L_\mathrm{X}$=7.7$\times$10$^{33}$~erg~s$^{-1}$ and
$n_\mathrm{e}\gtrsim$0.4~cm$^{-3}$.

We have recently obtained {\it XMM-Newton} observations of this nebula
and have confirmed that the distribution of the X-ray-emitting gas
presents three maxima: two associated to the caps and an extra
spatially correlated to the northwest blowout as suggested by the {\it
Chandra} observations (see Fig.~1, top right panel). The global
physical properties of the X-ray emission in NGC\,6888 are very
similar as those obtained for the {\it Chandra} observations, but due
to the superior spectral capabilities of the EPIC cameras we are able
to detect spectral variations (temperature and nitrogen abundance)
when studying in detail different regions within the nebula (Toal\'{a}
et al. in prep.). It is probable that the variations in nitrogen
abundance might be due to different mixing efficiencies in different
regions of the nebula as suggested by the possible interaction with a
cold filament as seen in infrared wavelegths \citep[see discussion
by][]{Toala2014}.


\section{Remarks}

Even though these three WR nebulae might have formed as the result of
different stellar evolutionary paths, they all exhibit diffuse X-ray
emission with soft temperatures. The main reason of this is now
understood: mixing of nebular material into the hot bubble by
instabilities and/or thermal conduction, but the contribution of each
effect has not been critically assessed so far. It would be also
interesting to test numerically the effect of cooling due to the dust
present in the nebula \citep{Toala2015WISE}.

It is worth noting that the WR nebulae reported to harbor diffuse
X-ray emission have very similar central stars: WN4--6 type stars with
terminal wind velocities of $v_{\infty}\approx$1700~km~s$^{-1}$ and
mass-loss rates of
$\dot{M}$=2--5$\times$10$^{-5}$~M$_{\odot}$~yr$^{-1}$
\citep{Hamann2006}. Those that do not exhibit diffuse X-ray emission
have WN8h stars (e.g., WR\,16 and WR\,40) with lower stellar wind
velocities ($v_{\infty}$=650~km~s$^{-1}$) but similar mass-loss rates.
In addition, our recently obtained {\it XMM-Newton} observations of
the WR nebula NGC\,3199 around WR18 also follow this trend (Toal\'{a}
et al. in prep.): diffuse X-ray emission is detected in a WR nebula
around a WN4-type star ($v_{\infty}\approx$1700~km~s$^{-1}$,
$\dot{M}$=3$\times$10$^{-5}$~M$_{\odot}$~yr$^{-1}$). A detailed
analysis on these observations is on-going in order to assess the
physical properties of the diffuse X-ray emission and put it in
context with S\,308, NGC\,2359 and NGC\,6888.


\bibliographystyle{aa} 
\bibliography{myarticle}

\end{multicols}

\end{contribution}


\end{document}